# Efficient 4.42 µm Raman laser based on hollow-core silica fiber

MAXIM S. ASTAPOVICH,[1] ANTON N. KOLYADIN,[1] ALEXEY V. GLADYSHEV,[1,*]
ALEXEY F. KOSOLAPOV,[1] ANDREY D. PRYAMIKOV,[1]
MAXIM M. KHUDYAKOV,[1] MIKHAIL E. LIKHACHEV,[1] AND IGOR A. BUFETOV[1]

[1]*Fiber Optics Research Center, Russian Academy of Sciences, 38 Vavilov St., 119333 Moscow, Russia*
*\*alexglad@fo.gpi.ru*

**Abstract:** In this paper we consider mid-infrared Raman lasers based on gas-filled hollow-core silica fibers and provide theoretical and experimental analysis of factors that limit the efficiency and output power of these lasers. As a result, we realized an efficient ns-pulsed 4.42 µm Raman laser based on an $^1H_2$-filled revolver silica fiber. Quantum efficiency as high as 36 % is achieved, and output average power as high as 250 mW is demonstrated. The possibilities of further improving the laser efficiency are discussed.

## References and links

1.  B. M. Walsh, H. R. Lee, and N. P. Barnes, "Mid infrared lasers for remote sensing applications," J. Lumin. **169**, Part B, 400–405 (2016).
2.  B. Wedel, and M. Funck, "Industrial Fiber Beam Delivery System for Ultrafast Lasers," Laser Tech. J. **13**(4), 42–44 (2016).
3.  A. D. Pryamikov, A. S. Biriukov, A. F. Kosolapov, V. G. Plotnichenko, S. L. Semjonov, and E. M. Dianov, "Demonstration of a waveguide regime for a silica hollow-core microstructured optical fiber with a negative curvature of the core boundary in the spectral region > 3.5 µm," Opt. Express **19**(2), 1441–1448 (2011).
4.  F. Yu, W. J. Wadsworth, and J. C. Knight, "Low loss silica hollow core fibers for 3–4 µm spectral region," Opt. Express **20**(10), 11153 – 11158 (2012).
5.  A. N. Kolyadin, A. F. Kosolapov, A. D. Pryamikov, A. S. Biriukov, V. G. Plotnichenko and E. M. Dianov, "Light transmission in negative curvature hollow core fiber in extremely high material loss region," Opt. Express **21**(8), 9514–9519 (2013).
6.  F. Yu and J. C. Knight, "Spectral attenuation limits of silica hollow core negative curvature fiber," Opt. Express **21**(18), 21466–21471 (2013).
7.  F. Yu and J. C. Knight, "Negative Curvature Hollow-Core Optical Fiber," IEEE J. Sel. Topics Quantum Electron., **22**(2), 4400610 (2016)
8.  A. M. Jones, A. V. Nampoothiri, A. Ratanavis, T. Fiedler, N. V. Wheeler, F. Couny, R. Kadel, F. Benabid, B. R. Washburn, K. L. Corwin, and W. Rudolph, "Mid-infrared gas filled photonic crystal fiber laser based on population inversion," Opt. Express **19**(3), 2309-2316 (2011).
9.  A. V. Nampoothiri, A. M. Jones, C. Fourcade-Dutin, C. Mao, N. Dadashzadeh, B. Baumgart, Y.Y. Wang, M. Alharbi, T. Bradley, N. Campbell, F. Benabid, B. R. Washburn, K. L. Corwin, and W. Rudolph, "Hollow-core Optical Fiber Gas Lasers (HOFGLAS): a review [Invited]," Opt. Mater. Express **2**(7), 948–961 (2012).
10. N. Dadashzadeh, M. P. Thirugnanasambandam, N. K. Weerasinghe, B. Debord, M. Chafer, F. Gerome, F. Benabid, B. R. Washburn, and K. L. Corwin, "Near diffraction-limited performance of an OPA pumped acetylene-filled hollow-core fiber laser in the mid-IR," Opt. Express **25**(12), 13351-13358 (2017).
11. M. Xu, F. Yu, and J. Knight, "Mid-infrared 1 W hollow-core fiber gas laser source," Opt. Lett., **42**(20), 4055–4058 (2017).
12. A. V. Gladyshev, A. F. Kosolapov, M. M. Khudyakov, Yu. P. Yatsenko, A. K. Senatorov, A. N. Kolyadin, A. A. Krylov, V. G. Plotnichenko, M. E. Likhachev, I. A. Bufetov, E. M. Dianov, "Raman Generation in 2.9 - 3.5 µm Spectral Range in Revolver Hollow-Core Silica Fiber Filled by $H_2/D_2$ Mixture," in *Conference on Lasers and Electro-Optics*, OSA Technical Digest (online) (Optical Society of America, 2017), paper STu1K.2.
13. E. B. Kryukova, V. G. Plotnichenko, and E. M. Dianov, "IR absorption spectra in high-purity silica glasses fabricated by different technologies," Proc. SPIE 4083, 71–80 (2000).
14. A. V. Gladyshev, A. F. Kosolapov, M. M. Khudyakov, Yu. P. Yatsenko, A. N. Kolyadin, A. A. Krylov, A. D. Pryamikov, A. S. Biriukov, M. E. Likhachev, I. A. Bufetov, E. M. Dianov, "4.4-µm Raman laser based on hollow-core silica fibre", Quantum Electron., **47**(5), 491–494 (2017).
15. I. A. Bufetov and E. M. Dianov, "A simple analytic model of a CW multicascade fibre Raman laser," Quantum Electron., **30**(10), 873–877 (2000).

16. M. Miyagi, "Bending losses in hollow and dielectric tube leaky waveguides," Appl. Opt., **20**(7), 1221–1229 (1981).
17. C. Headley and G. P. Agrawal, *Raman amplification in fiber optical communication systems* (Academic Press, 2005).
18. D. C. Hanna, D. J. Pointer, and D. J. Pratt, "Stimulated Raman Scattering of Picosecod Light Pulses in Hydrogen, Deuterium, and Methane", IEEE J. Quantum Electron., **22**(2), 332–336 (1986).
19. J. F. Reintjes, "Stimulated Raman and Brillouin scattering" in *Handbook of Laser Science and Technology*, Suppl. 2: Optical Materials (CRC Press, 1995), p. 334.
20. W. K. Bischel and G. Black, "Wavelength dependence of Raman scattering cross sections from 200–600 nm", AIP Conference Proceedings, **100**, 181–187 (1983).
21. A. V. Gladyshev, A. N. Kolyadin, A. F. Kosolapov, Yu. P. Yatsenko, A. D. Pryamikov, A. S. Biriukov, I. A. Bufetov and E. M. Dianov, "Low-threshold 1.9 µm Raman generation in microstructured hydrogen-filled hollow-core revolver fibre with nested capillaries," Laser Phys. **27**, 025101 (2017).
22. A. V. Gladyshev, A. N. Kolyadin, A. F. Kosolapov, Yu. P. Yatsenko, A. D. Pryamikov, A. S. Biriukov, I. A. Bufetov, E. M. Dianov, "Efficient 1.9-mm Raman generation in a hydrogen-filled hollow-core fibre," Quantum Electron., **45** (9), 807–812 (2015).
23. F. Poletti, "Nested antiresonant nodeless hollow core fiber," Opt. Express, **22**(20), 23807–23828 (2014).
24. Y. Chen, Z. Wang, B. Gu, F. Yu, and Q. Lu, "Achieving a 1.5 µm fiber gas Raman laser source with about 400 kW of peak power and a 6.3 GHz linewidth," Opt. Lett., 41(21), 5118–5121 (2016).

## 1. Introduction

Laser sources emitting in 3÷5 µm spectral range are required for various applications in biomedicine, gas analysis and material processing [1]. Fiber lasers are of a particular interest as they can provide excellent beam quality, alignment-free operation and compact design. One of the approaches for development of mid-infrared (mid-IR) fiber lasers is based on gas-filled hollow core fibers (HCF). Compared with solid core fibers for mid-IR, the HCFs can potentially withstand much higher peak power without damaging the fiber [2]. Moreover, due to strong localization of the optical field within a hollow core, HCFs can have low optical losses even in those spectral regions, where a cladding material has strong fundamental absorption. For example, mid-IR light guidance was demonstrated in the hollow core fibers that had the cladding made of silica glass [3–7].

Mid-IR lasers based on gas-filled hollow core silica fibers have been realized in several works, where radiation at wavelengths around $\lambda \sim 3$ µm was generated using population inversion [8–11] or stimulated Raman scattering (SRS) [12]. Radiation generation at longer wavelengths is a challenging problem as the material absorption of silica cladding in the 3÷5 µm spectral range grows dramatically from ~50 to ~50000 dB/m [13]. Nevertheless, due to an extremely low overlap of the optical mode with the cladding, revolver-type HCFs have recently enabled Raman generation at a wavelength as long as 4.42 µm [14]. So far, this wavelength is the longest to have been generated in gas-filled hollow-core silica fibers. However, the average power and quantum efficiency of the 4.42 µm Raman laser were limited to 30 mW and 15 %, respectively [14].

In this work we analyze factors that limit the efficiency of mid-IR nanosecond Raman lasers based on hydrogen-filled hollow-core silica fibers. First, we estimate the optimal diameter of the hollow core by using simple analytical models developed previously for 1) fiber Raman lasers [15] and 2) optical losses of bent hollow-core fibers [16]. Then, we realize a 4.42 µm Raman laser, and study the impact of optical losses and transient SRS effects on the laser efficiency. Finally, we demonstrate a simple way to avoid detrimental effects caused by a transient SRS regime in ns-pulsed Raman lasers, thus maximizing output power and efficiency.

## 2. Methods

To evaluate a fiber as an active medium for single-cascade Raman conversion, we used the parameter $P_F$ [15]:

$$P_F = \left(\sqrt{\frac{\alpha_P}{g_0}} + \sqrt{\frac{\alpha_S}{g_0}}\right)^2 \equiv \frac{A_{eff}}{g_R}\left(\sqrt{\alpha_P} + \sqrt{\alpha_S}\right)^2 \tag{1}$$

which can be considered as a figure of merit of a Raman fiber. This parameter takes into account the combined effect of several fiber properties: the optical losses ($\alpha_P$ and $\alpha_S$) at both pump and Stokes wavelength, respectively, and the Raman gain of the fiber ($g_0$), which can be alternatively expressed in terms of the Raman gain of a medium ($g_R$) and the effective area ($A_{eff}$) for the Raman conversion in this fiber [17]. Being introduced in [15] for solid-core fiber Raman lasers, the parameter $P_F$ can be equally applied to the case of gas-filled hollow-core fibers. The parameter $P_F$ has dimensionality of power and has a meaning of pump power that is required to reach the SRS threshold in the fiber placed into an idealized resonator [15]. Thus, the lower the value of $P_F$ is, the more the fiber is appropriate for Raman conversion.

To estimate the optimal diameter of the hollow core ($D_{opt}$) for the 1.56 → 4.42 μm Raman conversion, the parameter $P_F$ was calculated as a function of the core diameter ($d_c$). The calculation was based on a simplified model that describes optical losses of a hollow-core fiber and includes bend-induced optical losses [16]. The model takes into account the interference effects of the optical waves reflected in the thin glass film at the core-cladding boundary, but does not consider the negative curvature effects of this boundary. For revolver-type HCFs this simplified model overestimates the absolute values of optical losses, thus overestimating the absolute value of $P_F$. However, relative values are sufficient to estimate the optimal diameter of the hollow core because this diameter can be identified by a local minimum of the function $P_F = f(d_c)$.

The value of the Raman gain $g_R$ for the 1.56 → 4.42 μm conversion in molecular hydrogen is not available in the literature. For hydrogen at pressures above ~ 10 atm we calculated the value of a steady-state vibrational Raman gain to be $g_R$ = 0.43 cm/GW. The calculation was based on a well-known expression for $g_R$ (see, e.g. [18]) and on published data on the vibrational Raman linewidth and scattering cross-section of $^1H_2$ [19, 20]. This method of calculation was previously verified in [21, 22], where $g_R \approx 1$ cm/GW was calculated for the 1.06 → 1.91 μm vibrational SRS in accordance with experimental data.

The optical loss spectrum of the fiber was calculated using COMSOL Multiphysics software and taking into account the material absorption of silica glass. The measurement of optical losses at wavelengths below 2 μm was carried out by a cut-back technique using radiation of a supercontinuum source (Fianium) at the input to the fiber. The accuracy of the loss measurement was limited to 0.03 dB/m as a result of power instabilities of the source and a relatively short length of the fiber cut section ($\Delta L$= 20 m). To measure optical losses at the Stokes wavelength ($\lambda$ = 4.42 μm), we made use of the previously developed 4.42 μm low-power silica fiber Raman laser [14]. Radiation of that laser was coupled to the revolver fiber, and the fiber transmission ($T$) was measured as a function of the fiber length ($L$), while the length was successively shortened in ~ 40 cm steps. The value of optical losses ($\alpha$) was then determined according to the relation $\alpha = -\frac{d}{dL}\left(10 \log_{10}(T)\right)$. Note, the Raman laser generated the 4.42 μm radiation in a single transversal mode, thus enabling reliable measurements of optical losses of the fundamental mode at 4.42 μm.

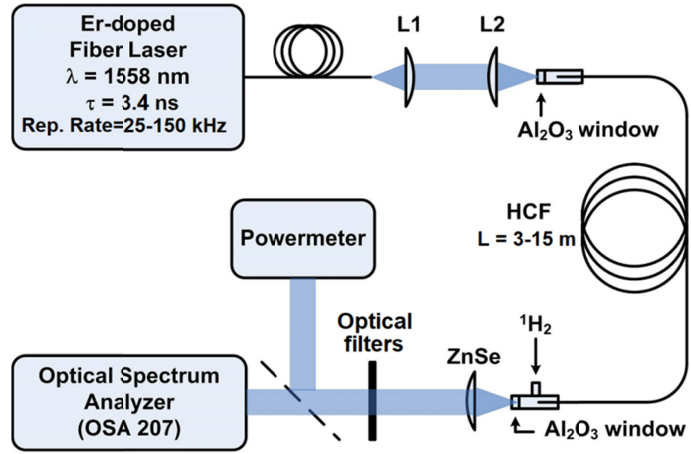

Fig. 1. The scheme of experimental setup.

The experimental setup shown in Fig. 1 was used to study the factors that limit the efficiency of mid-IR nanosecond gas fiber Raman lasers. A hydrogen-filled revolver fiber was pumped through lenses L1 and L2 by an Er-doped fiber laser at the 1.56 μm wavelength. The pump laser emitted 3.4 ns pulses with an output average power of 2.4 W. The repetition rate of the pulses can be varied in the 25–150 kHz range. The coupling efficiency of the pump power into the HCF was as high as 80 %. Radiation at the HCF output was collimated by a ZnSe lens and passed through optional optical filters. Optical spectra and power were then measured by an optical spectrum analyzer (OSA207, Thorlabs) and a powermeter (3A-P-SH-V1, Ophir). The beam profile and the temporal pulse shape at the Stokes wavelength were observed by a mid-infrared camera (Pyrocam IV, Ophir) and a photoelectromagnetic (HgCd)Te detector (PEM, VIGO System S.A.).

To analyze vibrational SRS in a revolver fiber filled by molecular hydrogen $^1H_2$, coupled wave equations were numerically solved in COMSOL Multiphysics. Only two waves were taken into account: the pump wave at 1.56 μm and the Stokes wave at 4.42 μm. Both waves were assumed to propagate in a fundamental transversal mode of the revolver fiber. The measured pulse shape of the pump pulses was used in simulations. The value of the Raman gain was taken to have the steady-state value of 0.43 cm/GW.

## 3.  Results and discussion

The optimal diameter $D_{opt}$ of the hollow core can be estimated using Fig. 2, where the calculated parameter $P_F$ is shown as a function of the core diameter $d_c$ for several values of bending radius $R$. Having a meaning of threshold pump power of an idealized Raman laser (see [15]), the parameter $P_F$ reaches a minimum when the diameter of a hollow core has the optimal value. For each value of the bending radius, there is a single optimal core diameter $D_{opt}$ (Fig. 2a). When the core diameter deviates towards lower values ($d_c < D_{opt}$) the Raman threshold increases due to rapid growth of the optical losses even if the fiber is assumed to be straight ($\alpha \sim 1/d_c^3$ [16, 23]). In the opposite case ($d_c > D_{opt}$) the Raman threshold increases as well (Fig. 2a) due to the growth of bend-induced losses [16] in combination with reduced intensity of the pump wave.

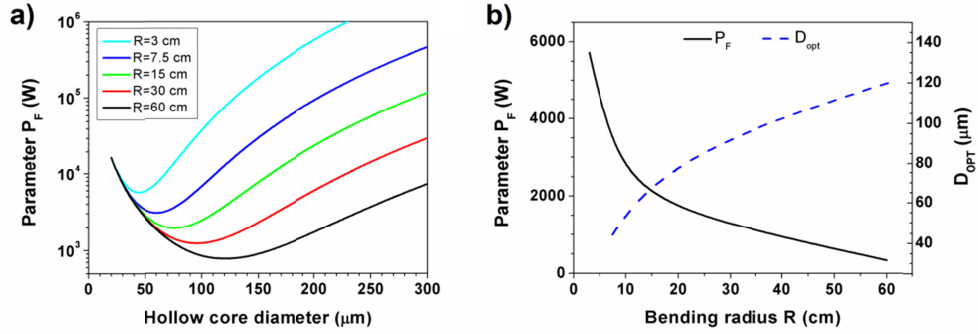

Fig. 2. (a) The parameter $P_F$ as a function of the hollow core diameter $d_c$ at several fixed values of the bending radius $R$ of the fiber. The curves were calculated for 1.56 to 4.42 µm Raman conversion in a hollow-core fiber filled by molecular hydrogen at pressure of 20 bar ($g_R$ = 0.43 cm/GW). Optical losses, required to calculate $P_F$, were simulated according to model developed in [16]. (b) The minimal value of $P_F$ (solid curve) and the optimal core diameter $D_{opt}$ (dashed curve) as a function of the bending radius of the fiber.

Represented by the parameter $P_F$, the minimal value of threshold pump power monotonically decreases with the bending radius of a hollow-core fiber (Fig. 2b, solid curve). However, at the same time the laser loses its compact size, which is one of the important advantages of fiber lasers. In this work the bending radius $R$ = 15 cm was taken as an appropriate value for practical devices. Thus, the optimal diameter of the hollow core was estimated to be about $D_{opt}$ = 75 µm (Fig. 2b, dashed curve).

For experimental investigation a revolver-type HCF with optimal core diameter of 75 µm was fabricated (Fig. 3a, inset). The calculated fundamental mode field diameter was 55 µm and the cladding was formed by ten silica (F300) capillaries with a 21.7 µm inner diameter and 1.15 µm wall thickness.

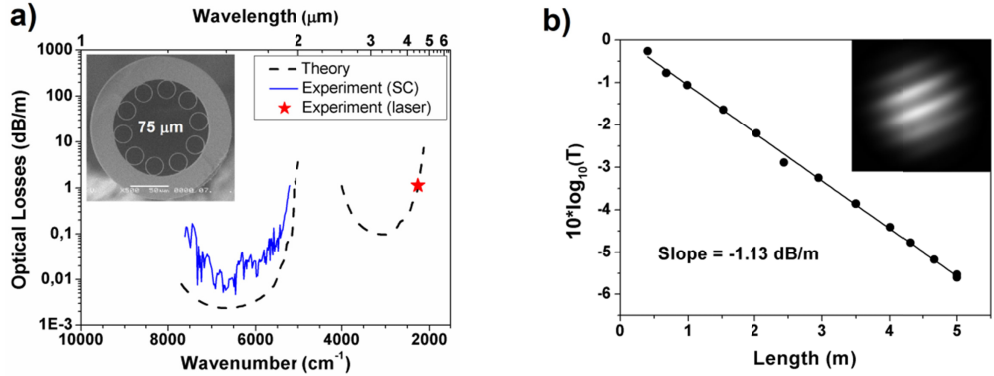

Fig. 3. (a) The spectrum of optical losses calculated for the revolver fiber (black dashed curves). The losses measured using supercontinuum source (blue solid curve) or using single-mode 4.42 µm laser (red star) are also shown. SEM image of the revolver fiber cross section is shown in the inset. (b) Transmission measured at $\lambda$ = 4.42 µm as a function of the revolver fiber length. The mode field distribution (inset) corresponds to a fundamental mode modulated by interference on the entrance window of the beam profiling camera.

The calculated loss spectrum of the fiber is shown in Fig. 3a (dashed curves). Measured optical losses (Fig. 3a, blue solid curve, red star) correlate well with calculations results. To analyze the Raman conversion we are particularly interested in loss values for the pump ($\lambda$ = 1.56 µm) and the Stokes ($\lambda$ = 4.42 µm) waves. At the pump wavelength the measured attenuation has a value of 0.03 dB/m, which in fact is equal to the accuracy of the measurements. To determine optical losses at the Stokes wavelength, fiber transmission ($T$) at this wavelength was measured as a function of the fiber length (Fig. 3b). As a result, optical losses at $\lambda = 4.42$ μm were found to be as low as 1.13 dB/m. Note, the material absorption of silica cladding at the 4.42 μm wavelength amounts to ~ 4000 dB/m [13].

Using the measured values of optical losses, we evaluated the revolver fiber as an active medium for mid-IR gas Raman lasers. For this purpose, vibrational SRS was numerically modeled assuming the hollow core was filled by molecular hydrogen $^1H_2$ at the 20 atm pressure and at room temperature. Parameters of the pump radiation were chosen to fit experimental conditions (wavelength $\lambda = 1.56$ μm, pulse duration $\tau = 3.4$ ns, repetition rate $f = 25$ kHz). The average power of the Stokes wave ($\lambda = 4.42$ μm) calculated as a function of fiber length is plotted in Fig. 4 (solid curve) for the case when 15 kW peak pump power is coupled to the fiber. One can see that the rapid growth of the Stokes power is completed within a few first meters of the fiber. At a longer length the Stokes power decreases monotonically due to optical losses in the fiber. This simulation shows that the optimal fiber length has a value about 3 – 3.5 m, which is much lower compared with the value of 15 m used in the previous work [14].

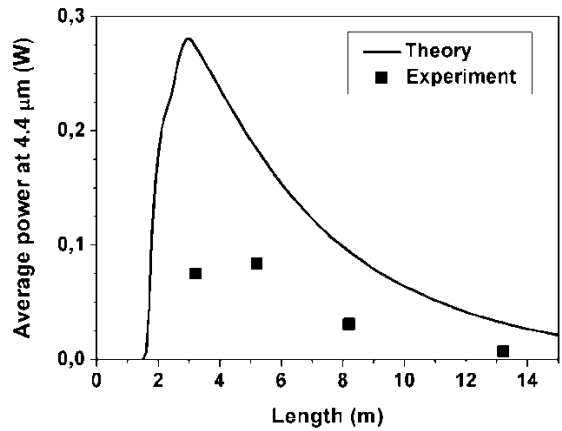

Fig. 4. Calculated (line) and measured (squares) average output power at $\lambda = 4.42$ μm is shown as a function of length of the Raman laser. Peak pump power coupled to the revolver fiber was 15 kW. Pump pulse duration and repetition rate were 3.4 ns and 25 kHz, respectively.

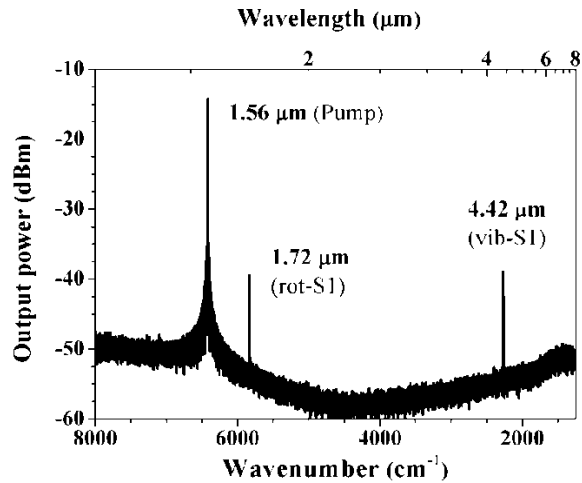

Fig. 5. Typical output spectrum of the Raman laser based on hydrogen-filled hollow core silica fiber.

To determine the optimal fiber length in the experiment, first, a 4.42 µm Raman laser was built using an intentionally long (L = 13.2 m) revolver fiber coiled with the radius of 15 cm. Then, the output power of the laser was measured as a function of the fiber length while the fiber was sequentially shortened. The characteristics of the 13.2-m-long laser were similar to those reported in [14]. The maximum average power at 4.42 µm was about 30 mW, corresponding to the quantum conversion efficiency of ~ 15 %. A typical spectrum measured at the output of the Raman laser shows a competition of two nonlinear processes (Fig. 5). The Stokes wave at λ = 4.42 µm was generated due to vibrational SRS on Q(1) transition of hydrogen molecules $^1H_2$ ($\Omega_{Vib}$ = 4155 cm$^{-1}$). The spectral line at λ = 1.72 µm corresponds to the first Stokes wave of rotational SRS on $S_0(1)$ transition of $^1H_2$ molecules ($\Omega_{Rot}$ = 587 cm$^{-1}$). In this work, all the experiments on the Raman laser optimization were carried out in those regimes where vibrational SRS dominates.

The experimental optimization of the fiber length confirms general behavior of the Stokes power and justifies the value of optimal length (Fig. 4, squares). However, the measured power of the Stokes wave appeared to be ~ 3 times lower in comparison with the calculated Stokes power. This discrepancy could be a manifestation of a transient SRS regime, which is known to reduce the value of the Raman gain compared with the gain in steady-state SRS.

To verify the above assumption, let us compare the duration of pump pulses (τ = 3.4 ns) with the dephasing time of molecular vibrations ($T_2$). Typically, a transient SRS regime takes place when τ < 20$T_2$ [18]. The value of dephasing time is defined as $T_2$ = 1/(π$\Delta v_R$), where $\Delta v_R$ is the Raman linewidth (FWHM). The value of $\Delta v_R$ is, in turn, a function of gas density (ρ), and for vibrational Q(1) transition of $^1H_2$ molecules at room temperature it can be approximated as $\Delta v_R$ =( 309/ρ) + 51.8ρ, where ρ is measured in amagat and $\Delta v_R$ is in MHz [19]. In the case of hydrogen pressure of 20 atm we find that $T_2$ = 0.33 ns, thus τ ≈ 10$T_2$ and the condition for the transient regime is satisfied.

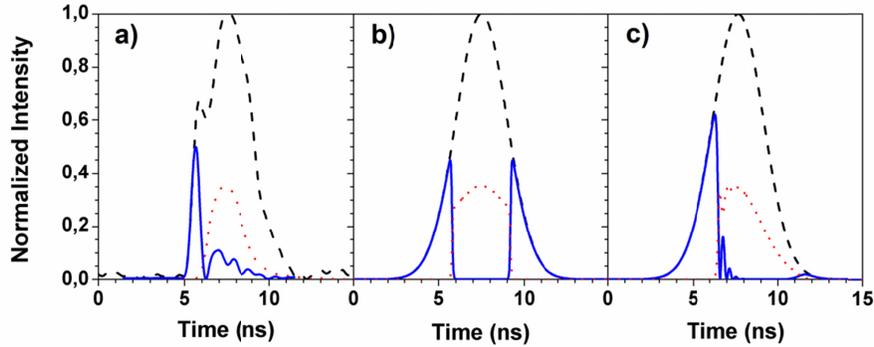

Fig. 6. (a) The shapes of pump (blue solid curve) and Stokes (red dotted curve) pulses measured at the output of the fiber Raman laser. The shape of pump pulse measured at the input to the fiber is also shown (black dashed curve). (b) Simulation results of the same dependences for the case of steady-state SRS. (c) Simulation results of the same dependences for the case of transient SRS. All the data correspond to the case when the peak pump power of 15 kW was coupled into 3.2-m-long revolver fiber filled by $^1H_2$ at pressure of 20 atm.

The evidence of the transient SRS regime was found by measuring the temporal shape of optical pulses at the output of the Raman laser (Fig. 6a). In the experiment the peak pump power of 15 kW was coupled into a 3.2-m-long revolver fiber filled by $^1H_2$ at a pressure of 20 atm. The pump pulse shape at the input to the fiber is shown in Fig. 6a by a black dashed curve. One can see that only a leading edge of the input pump pulse is detected at the output of the Raman laser (Fig. 6a, blue solid curve). This asymmetric shape of the output pulses at the pump wavelength should be compared with the results of numerical simulations, which were carried out for both steady-state (Fig. 6b) and transient (Fig. 6c) Raman scattering. A steady-state SRS regime is characterized by the symmetric shape of the residual pump pulses

(Fig. 6b, blue solid curve; see also [21]), whereas the asymmetry of output pump pulses is a signature of a transient SRS regime (Fig. 6c, blue solid curve).

The fact that the Raman linewidth depends on gas density provides a way to adjust the value of $T_2$ simply by changing gas pressure. In this way, a steady state SRS regime could be realized even if the pump pulses have a duration of a few ns. To verify this, we measured the output pulse energy of a 4.42 μm Raman laser while changing the pressure of hydrogen in the hollow core from 10 to 70 atm (Fig. 7). One can see that pulse energy increases linearly with hydrogen pressure, reaching saturation at a pressure of about ∼ 50 atm. The vertical dashed line corresponds to the pressure, at which $\tau = 20T_2$ for the case of the pump pulses used ($\tau = 3.4$ ns) in the experiment. This line splits the data into two ranges, where a transient ($\tau < 20T_2$) or steady state ($\tau > 20T_2$) SRS regime dominates as confirmed by experimental data. The obtained dependence (Fig. 7) proves that the detrimental effect of a transient SRS regime can be avoided in ns-pulsed gas-filled fiber Raman lasers by means of simple adjustment of gas pressure.

In the following, the Raman laser was studied in optimized conditions: the pressure of hydrogen was set to 50 atm and the length of the revolver fiber was set to 3.2 m. In addition, the average power of the Stokes component at $\lambda = 4.42$ μm was maximized by setting the repetition rate of the pump pulses to 50 kHz.

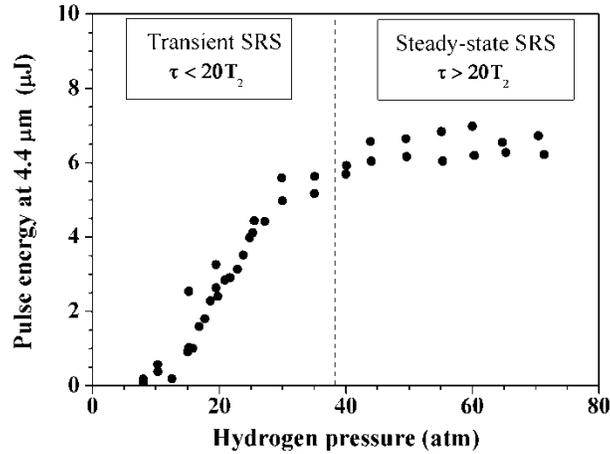

Fig. 7. Output pulse energy at $\lambda = 4.42$ μm as a function of $^1H_2$ pressure. Measurements carried out at different repetition rates reproduce the same dependence on pressure. Length of the fiber was 3.2 m.

The output power measured for each spectral component is shown in Fig. 8 as a function of average pump power coupled to the fiber. In spite of a large quantum defect for the $1.56 \rightarrow 4.42$ μm conversion, the average power as high as 250 mW was generated at $\lambda = 4.42$ μm when coupled pump power was 1.9 W (Fig. 8, red dots). In this regime the Stokes pulses had a duration of ∼2 ns and peak power of ∼2.5 kW. The power conversion efficiency of 13 % was achieved, corresponding to quantum conversion efficiency as high as 36%.

The measured (Fig. 8, dots) and calculated (Fig. 8, solid curves) power dependencies are in good agreement with each other. Some deviations of the measured power at $\lambda = 4.42$ μm from the calculated values can be attributed to the excitation of higher-order modes at the pump wavelength and, partially, to the occurrence of the rotational Stokes wave at $\lambda = 1.72$ μm.

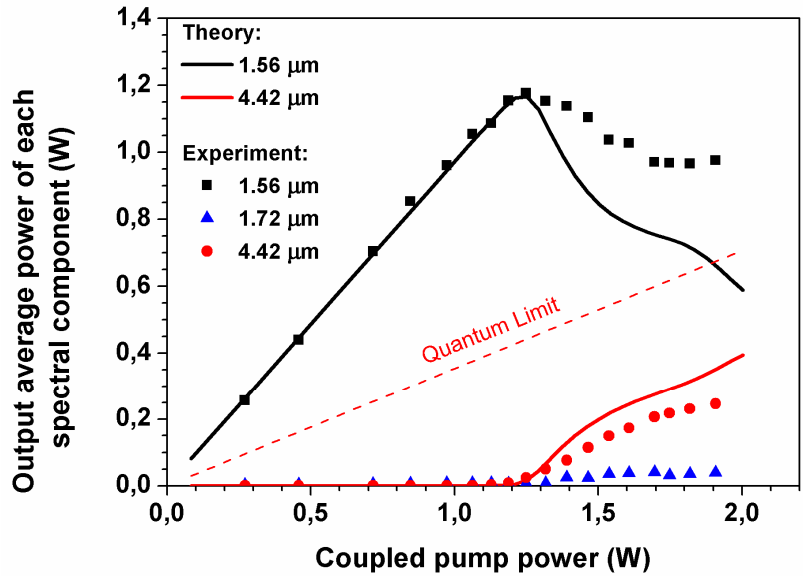

Fig. 8. The average output power of each spectral component of the Raman laser as a function of coupled average pump power. Measured data are shown for residual pump (black squares), vibrational Stokes at 4.42 μm (red dots) and rotational Stokes at 1.72 μm (blue triangles). Results of numerical calculations are also shown for vibrational Stokes (red solid curve) and residual pump (black solid curve). Red dashed line represents a quantum limit for generation at 4.42 μm. The output power is corrected for transmission of filters, ZnSe lens and $Al_2O_3$ output window.

The results of numerical simulations (Fig. 8, red solid curve) show that the maximum quantum efficiency of the Raman laser can be as high as 55 %. This value is mainly limited by optical losses at the Stokes wavelength ($\alpha = 1.13$ dB/m), whereas the losses at the pump wavelength ($\alpha \leq 0.03$ dB/m) have little effect on the performance of the Raman laser due to a relatively short length of the fiber used (3.2 m). Another parameter, which limits the efficiency, is the maximum peak power of the pump laser. The higher the power available to pump the Raman laser, the shorter the fiber length is required, thus reducing the detrimental effects of optical losses.

In our experiments the maximum peak power of the pump source (~ 30 kW) was limited by nonlinear effects in a solid core of an Er-doped fiber laser. However, this value can be significantly increased if a hollow-core fiber is used to generate the 1.56 μm radiation. In this way, a fiber laser emitting ns-pulses with the peak power of 400 kW at $\lambda = 1.55$ μm was recently demonstrated [24]. Being implemented as a pump source for gas-filled revolver fibers, such lasers could significantly improve the efficiency of the 1.56 → 4.42 μm conversion. Moreover, an efficient two cascade 1.06 → 1.55 → 4.42 μm Raman conversion should be possible in a single piece of a revolver fiber filled by a mixture of ethane and hydrogen gases. Based on this approach, one can expect a development of mid-IR silica fiber Raman lasers that can generate nanosecond pulses with peak power approaching ~100 kW in the 3–5 μm spectral range.

### 4. Conclusions

To conclude, we have demonstrated a revolver silica fiber with measured optical losses as low as 1.13 dB/m at a wavelength of 4.42 μm. Filled by molecular hydrogen, this fiber enabled realization of an efficient gas fiber Raman laser at 4.42 μm. Pumped by 3.4 ns pulses of an Er-doped fiber laser at 1.56 μm, the Raman laser generated output average power as high as 250 mW at 4.42 μm with quantum conversion efficiency as high as 36 %. In this regime the Stokes pulses had a duration of ~2 ns and the peak power of ~2.5 kW. It was shown that the detrimental effects caused by the transient regime in ns-pulsed Raman lasers can be avoided by means of simple adjustment of gas pressure. We believe that the efficiency and peak power of mid-IR Raman lasers based on gas-filled revolver fibers can be further improved. Such lasers will find numerous applications in science, biomedicine and technology.

**Funding**

Russian Science Foundation (RSF) (№16-19-10513).

**Acknowledgments**

The authors would like to thank Dr. S. V. Muravyev for providing mid-infrared beam profiling camera. Also, the authors are grateful to Dr. Yu. M. Klimachev for providing mid-infrared fast photodetector.